\documentclass{article}
\usepackage{hyperref}
\usepackage{graphicx}
\usepackage{authblk}

\title{Challenges related to system-of-systems for greening and climate adaptation in smart cities.}

\newcommand{\orcidauthorA}{0000-0001-5693-5217} 
\newcommand{\orcidauthorB}{0000-0002-7696-0046} 

\providecommand{\keywords}[1]
{
  \small
  \textbf{\textit{Keywords---}} #1
}

\author[$\dagger$]{Sarah Brandt \orcidauthorA{}}
\author[$\dagger$]{Julien Siebert \orcidauthorB{}}

\affil[$\dagger$]{Fraunhofer Institute for Experimental Software Engineering (IESE); firstname.lastname@iese.fraunhofer.de}

\begin{document}
\maketitle


\begin{abstract}

This paper presents the results of interviews conducted as part of the DYNASOS project. The objective was to collect challenges related to the design, implementation and management of system-of-systems (SoS) in the context of climate adaptation and greening of smart cities. 23 individuals from cities, academia, and industry were interviewed between March and May 2022 and 57 distinct challenges were collected and analyzed. Our results show that while technical issues (such as interoperability or data acquisition) persist, non-technical issues are the main obstacles. Difficulties in information sharing, effective communication, and synchronization between different actors are the most important challenges.

\end{abstract}

\section{Introduction}
In this paper, we present and discuss the results of interviews done during the DYNASOS\footnote{\url{https://dynasos.de/}} projects for the field of smart cities, specifically focusing on the use cases of climate adaptation and greening. The project DYNASOS, funded by the German Federal Ministry for Education and Research (BMBF), has the goal of understanding the current trends in software engineering research and to propose a roadmap for future research funding \cite{Adler2022}. In that context, six application domains were chosen. The smart city domain was one of them. Indeed, the implementation of smart city strategies requires the setting up of digital infrastructure allowing several systems (sensors, database, applications) to interact. The goal is to provide both a global overview of the city services and the possibility to host and enable an ecosystem of new services allowing a broader involvement of citizens (see for example services like \url{https://giessdenkiez.de/}, or platform such as \url{https://foodsharing.de/} or \url{https://mundraub.org/}).
The supporting and underlying digital infrastructure are sometimes described as digital ecosystems or platforms \cite{Koch2022} and belong to a larger group of what is called system-of-systems (SoS) \cite{Gorod2008}.

In the context of the project DYNASOS, we performed literature reviews and conducted interviews and workshops with representatives from industry and academia from the smart city domain. We specifically focused on the use case of climate adaptation and greening. Our first goal was to collect and organize challenges from the point of view of stakeholders involved at the level of city planning. The second goal is to understand how these challenges relate to Software and SoS Engineering.

This article is structured as follows. Section \ref{section-related-work} provides relevant definitions and an overview of the challenges mentioned in the literature regarding the two areas of SoS and smart cities. Section \ref{section-materials-methods} and \ref{section-results} presents respectively the methodology followed and the obtained results. Section \ref{section-discussion} presents a discussion of the results and section \ref{section-conclusions} the conclusion.

\section{Definitions and Related work}\label{section-related-work}

\subsection{System-of-systems (SoS)}
Over time, the definition of a system-of-systems (SoS) has evolved \cite{Gorod2008}, but today most studies in that domain agree on the following definition: a system-of-systems combines multiple systems to accomplish a task that none of the systems can accomplish alone, with each constituent system maintaining its own management, goals, and resources while coordinating and adapting within the SoS to achieve the overall goals of the SoS (when these are explicitly defined).
Since this first definition is relatively abstract, SoS scholars have supplemented it with a set of characteristics that help refine what a SoS is \cite{Maier1998,Delaurentis2005,Boardman2006}. \textit{Autonomy} - A constituent system has the ability to make independent decisions. This includes management and operational independence \cite{Maier1998}. This means that the different systems taking part in the SoS are developed and operated by different organizations but still cooperate together. \textit{Heterogeneity, geographic distribution, and connectivity} - Constituent systems might be different in nature (e.g., sensors, actuators, software, etc.), with different dynamics operating at different time scales. They are generally not physically located in the same place, but are interconnected (i.e., they exchange information). \textit{Belonging} - Constituent systems have the right and ability to choose to belong to a SoS. \textit{Evolution and Emergence} - The infrastructure, functionality, and objectives of the SoS may change over time and depend on the evolution of its constituent systems. The overall behavior of the SoS is an emergent property that depends on the interaction of its constituent systems. One type of SoS implementation that is commonplace today are digital ecosystems where constituent systems take the form of services (or app) available, for example, on a marketplace \cite{Koch2022} (a classifications of the different types of SoS have been provided in \cite{Maier1998,Dahmann2008}).

Developing such systems presents a variety of engineering challenges. The literature on SoS \cite{Dridi2020,Henson2013,Dogan2013} and more recently on digital ecosystems \cite{Koch2022} provide insight about challenges from the software and SoS engineering point of view. These challenges can be summarized and organized as follows: challenges on the level of SoS characteristics and quality attributes (e.g., safety, availability, interoperability, compatibility, etc.), challenges on the level of management and oversight (e.g., stakeholder heterogeneity, responsibilities and leadership, etc.), challenges on the level of design and implementation \cite{Dridi2020,Henson2013,Dogan2013}.

\subsection{SoS and Smart-city}
Increasing urbanization and resource consumption in cities bring their own set of sustainability challenges. Smart cities attempt to address these challenges through new technologies and by connecting information from people and all elements of the city. The underlying assumption is that by using data and information technologies from various systems, smart cities are able to provide efficient services, control and process optimization. \cite{Albino2015, Monstadt2019,Westraadt2018}.

The increasing digitization of cities enable the collection of large amounts of information. This can be used for monitoring and analysis, and even to develop new applications and systems to meet the city's needs. Smart city solutions are already being used in various fields such as building management, transportation, infrastructure, health, public safety, services and governance \cite{Westraadt2018,Daneva2018}. Cities are by nature complex systems in which several actors, systems and domains interact together. The feedback loops that occur can create emergent phenomena that are challenging to control (see for instance \cite{Lenk2020}). One of the goals of smart city solutions is, if not to control the complexity of cities, at least to make it tangible. However, the interaction and connection of different systems in cities is difficult to achieve, making the implementation of smart city projects and the desired collaboration between different stakeholders slow and tedious \cite{Prasetyo2020}.

The SoS approach has been explored in the field of smart cities. \cite{Cavalcante2016} provide an example of a traffic SoS in which several systems interact with each other to provide an efficient and intelligent traffic system. The goal is to monitor, control and optimize the traffic using several sensors (number of cars, air pollution, etc.) and real-time data (traffic lights, tolls, etc). Adjustments in the traffic light system and route optimization can help to prevent traffic jams and to reduce air pollution. The authors in \cite{Axelsson2018} described several smart technologies that are in use in Singapore (road tolls, public transportation service, operations centers for traffic surveillance, air quality monitoring, etc.), as well as the communication networks between all kind of actors and stakeholders (including people mobility and transportation goods). The authors pointed out that there is a lack of knowledge how to connect the different systems and domains. Another example is provided in \cite{Elnashai2021} and focuses on sustainable cities by adapting the transportation within the city. The idea considered the use of autonomous components and real time adaptation to drive energy efficiency. The vision is a  completely autonomous city with autonomous interacting systems. The authors in \cite{Payne2020} applied a SoS thinking approach to develop a governance dashboard for smart streetlights applying a SoS framework developed in \cite{Lee2019}. For a successful dashboard, the authors mentioned that efficient communicating information and improving data accuracy are necessary. 

Finally, \cite{Daneva2018} establishes a list of requirements for what a smart city should provide. This includes end-to-end experience, architecture, security and privacy and infrastructure requirements. The authors mentioned specific requirements like the fact that data collection has to be possible from multiple sources, need to be manageable, transformable and storable for specific applications. Different systems, technologies, and services must be integrable through standardization and interoperability, so that different domains of the city, and their systems and data, can be used within a SoS.

\section{Materials and Methods}\label{section-materials-methods}

A purposive sampling approach was conducted, including stakeholders from cities, academia and industry. The objective was to identify people working in the smart city field who are interested in the following topics: sustainability and resilience, climate change adaptation and protection, and blue-green infrastructures and nature based systems. Therefore, the sampling was carried out in the following areas: management and development of smart cities, sustainability projects in cities and applied sciences, geographic information systems, climatology and meteorology, information technology, hydrology and agriculture.
All of the interviewees were working on specific tasks related to smart cities and technologies for sustainability on a daily basis, providing comprehensive coverage of smart city challenges and a variety of perspectives.
Their expertise has allowed for a better understanding of the decision-making processes of cities and the complexity of the different areas to be taken into account.

In total, we conducted 21 semi-structured interviews from march to may 2022, with two interviewers for each interview. The content was captured through written notes and recordings. Recordings were used only with the consent of the interviewee. The duration of each interview was 60 minutes.

The interviews were driven by the following guiding questions:
\begin{enumerate}
    \item What are the current challenges and research questions in the development of SoS in the domain of sustainability, climate adaptation and greening in smart cities?
    \item What are exemplary use cases to explain these challenges and research questions?
    \item What are representative example systems that can be used to make these use cases tangible?
\end{enumerate}

Since the system-of-systems approach was not necessarily known (at least by that name)\footnote{Most interviewees acknowledged that their day-to-day projects are indeed related to interacting systems and have to deal with many different stakeholders, but very few were familiar with the concept of "system-of-systems" and what this can actually bring them.} by the interviewees, we first created a mock up ("vision") of what a system-of-systems approach to managing green and blue infrastructure in a city might look like.
The interviews were then conducted in 4 parts. The first part was devoted to explaining the purpose of the interviews and the context (DYNASOS project). The second part was used to introduce the interviewees and interviewers and to collect demographic data. In the third part, we presented our "vision" of a SoS for smart and we collected comments and feedback. And finally, the last part was devoted to identifying current challenges, use cases and exemplary systems. In summary, we collected the following information:

\begin{itemize}
 \item Demographics.
 \item Feedback about the presented vision, use case, and motivation.
 \item Existing systems, actors, stakeholders.
 \item Current challenges.
 \item Existing and possible SoS solutions.
 \item Further applications of SoS.
\end{itemize}


\section{Results}\label{section-results}

\subsection{Demographics}
\begin{table}[!htbp]
 \centering
 \begin{tabular}{|c|p{0.75\textwidth}|}
 \hline
 Country & Germany (19), Austria (2), Estonia (1), Singapore (1) \\
 \hline
 Work context & City (6), University (5), Applied research (6), Private company (6) \\
 \hline
 Role & Project leader (8), Manager (6), Researcher (4), Professor (3), CEO (1), Student (1)\\
 \hline
 \end{tabular}
 \caption{Summary of the interviewees demographics. The number of interviewees for each category is shown in parentheses.}
 \label{tab:demographics}
\end{table}

Table \ref{tab:demographics} summarizes the demographics information, the following section provides more details.

We performed interviews with a total of 23 experts from Germany (19), Austria (2), Estonia (1), and Singapore (1) from march 2022 until may 2022.

From a context perspective, six persons were working directly in cities (Köln (2), Kaiserslautern (1), Mönchengladbach (1), Freiburg (1), Iserlohn (1)); five persons in universities (Hochschule Weihenstephan-Triesdorf (1), Tallinn University of Technology (1), Leibniz Universität Hannover für Meteorologie und Klimatologie (1), Universität für Bodenkultur Wien (1), and NTU’s Asian School of the Environment (1)); six persons in applied research institutions (Fraunhofer IAO (1), Fraunhofer UMSICHT (2), Deutsches Institut für Urbanistik DIFU (3)); and six persons in private companies (IP SYSCON GmbH (2), berchtoldkrass space\&options (2), GRÜNSTATTGRAU (1), and Kompetenzzentrum Wasser Berlin KWB (1)).

From a position (role) perspective, eight persons were working as project leaders, six as managers, four as researchers, three as professor, one as CEO, and one interviewee was student at the time of the interviews.

Finally, we asked participants to name their main areas of interest and the topics they are working on (hereinafter referred to as " topics of interest "). Table \ref{tab:topics_of_interests} presents an overview of the interviewees' topics of interest. We first used this information to see if we had missed any important topics for our case study.


\begin{table}[!htbp]
    \centering
    \begin{tabular}{|p{0.35\textwidth}c|p{0.35\textwidth}c|}
        \hline
        Climate change adaptation and climate protection & 13 & e-Governance & 1\\
        \hline
        Digitization & 6 & System and IT Consulting & 1 \\
        \hline
        Urban transformation, planning and development & 5 & Data Governance & 1 \\
        \hline
        Green spaces and green infrastructures & 3 & Indoor Smart Farming & 1 \\
        \hline
        Water treatment & 2 & Biodiversity & 1 \\
        \hline
        Energy & 2 & Climate modeling & 1 \\
        \hline
        Civil engineering & 1 & Drones & 1\\
        \hline
        Open Data & 1 & Resilience & 1\\
        \hline
        Sustainability & 1 & Environmental justice & 1 \\
        \hline
    \end{tabular}
    \caption{Overview of the interviewees principal topics of interests as they mentioned them, ranked by order. The number corresponds to the number of persons mentioning this specific topic. Note that several topics have usually be mentioned by a each interviewee and therefore the numbers do not sum up to 23 (the total of persons interviewed).}
    \label{tab:topics_of_interests}
\end{table}

\subsection{Challenges}
In total, we collected a set of 57 distinct challenges that we organized into 8 categories: data, people, system and software, system-of-system (SoS), work structure and processes, resource conflicts, complexity, and legal aspects. Table \ref{tab:challenges_categories} provide a short description of each category and the number of distinct challenge for each category.

\begin{figure}
    \centering
    \includegraphics[width=0.75\textwidth]{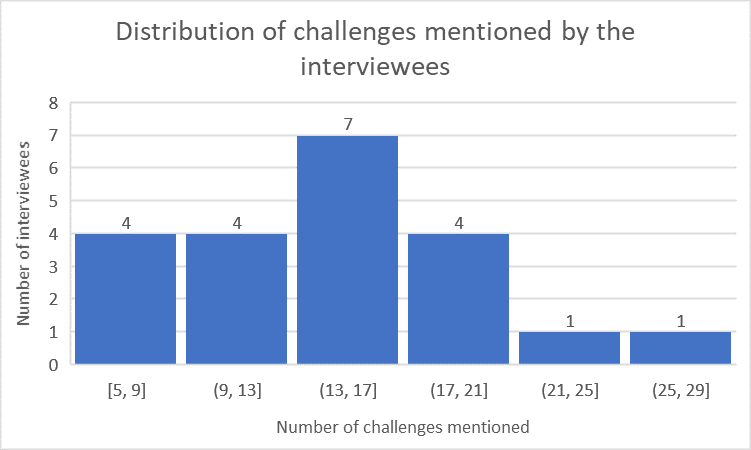}
    \caption{Distribution of the number of challenges mentioned by each interviewee. Most of the interviewees mentioned between 13 and 16 challenges (the median is 15). The minimum number of challenges mentioned by a single interviewee is 5 and the maximum is 27.}
    \label{fig:distrib_challenges_interviewees}
\end{figure}

Note that not all interviewees mention the same number of challenges. For each interviewees, we counted how many distinct challenges they mentioned. The resulting distribution is shown in figure \ref{fig:distrib_challenges_interviewees}.

We found that many of the challenges mentioned belonged to the categories "data", "people", "systems and software", "SoS", and "structure and work processes" (see table \ref{tab:challenges_categories}). This can be explained by the fact that most of the interviewees were from cities or worked closely with cities and were involved in smart city projects. Therefore, the number of distinct challenges identified in these categories was high in relation to resource conflicts, complexity, and legal aspects. Although the interviewees had no or very little contact with the SoS engineering field, they directly recognized the similarity between the SoS approach and their current projects. In addition, interviewees did not directly refer to SoS when mentioning their challenges, we classified them as "SoS" later.

\begin{figure}
    \centering
    \includegraphics[width=0.75\textwidth]{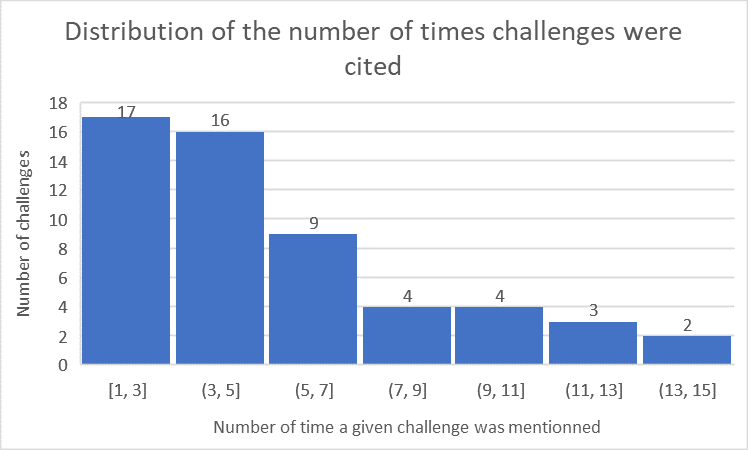}
    \caption{Distribution of the number of times a given challenge was mentioned by the interviewees. Half of the challenges were mentioned less than 5 times by all the interviewees (the median is 5, $Q1=2,5$ and $Q3=7$). Two challenges were cited each by 14 interviewees (max = 14). These two challenges concerns problems due to data silos and the communication between different actors.}
    \label{fig:distrib_importance_challenges}
\end{figure}

In order to estimate the relative importance of the challenges we counted how many times a given challenge was mentioned by the interviewees. The resulting distribution is shown in figure \ref{fig:distrib_importance_challenges}. Table \ref{tab:challenges_most_cited} presents the challenges that were most cited among the interviewees. For each category, tables \ref{tab:challenges_work_structure}, \ref{tab:challenges_data}, \ref{tab:challenges_people}, \ref{tab:challenges_software}, \ref{tab:challenges_sos}, and \ref{tab:challenges_other} provide the detail of all challenges extracted.

Among the most frequently mentioned challenges (table \ref{tab:challenges_most_cited}), we note first of all a predominance of challenges in the "data", "people" and "structure and work processes" categories (with 3, 3 and 2 challenges among the most important, respectively). We also find that although the categories "complexity", "resource conflicts" and "legal aspects" have fewer distinct challenges, they are relatively well ranked. Finally, we find that challenges in the "system and software" and "SoS" categories were ranked lower.

\begin{table}[!hbtp]
    \centering
    \begin{tabular}{|p{3cm}|p{7cm}|p{1,7cm}|}
        \hline
        \textbf{Category} &  \textbf{Description} & \textbf{Distinct challenges} \\
        \hline
        Data & Challenges related to the data lifecycle (e.g., acquisition, exchange, preparation, analysis, quality) & 12 \\
        \hline
        People & Challenges related to people (e.g., communication, skills, staffing, personal and political relations between departments) & 11 \\
        \hline
        System and software & Challenges related to system and software (e.g., legacy systems, missing features, user-friendliness) & 11 \\
        \hline
        SoS & Challenges related to the system-of-systems approach (e.g., distributed responsibility, security, impact) & 8 \\
        \hline
        Work structure and processes & Challenges related to the current work structure (e.g., agility, incentive to collaborate, transparency, processes incompatibility) & 8 \\
        \hline
        Resource conflicts & Challenges related to resource conflicts in cities (e.g., land use, costs) & 3 \\
        \hline
        Complexity & Challenges related to complex systems and emergent phenomena (e.g., foreseeabilty, non-linear feedback loops)  & 2 \\
        \hline
        Legal aspects & Challenges related to the different legal and regulatory aspects in cities (e.g., data privacy, regulations) & 2 \\
        \hline
    \end{tabular}
    \caption{Number of distinct challenges per category}
    \label{tab:challenges_categories}
\end{table}

\begin{table}[]
    \centering
    \begin{tabular}{|p{3cm}|p{6cm}|c|c|}
        \hline
        Category & Name & Mentions & Rank \\
        \hline
        Data & Data silos & 14 (87,5\%) & 1\\
        \hline
        People & Communication between different actors or stakeholders. & 14 (87,5\%) & 1\\
        \hline
        Work structure and processes & Stakeholders with their own plans, goals and interests, who are not willing to cooperate with each other. & 12 (75\%) & 3\\
        \hline
        Data & Data capture & 12 (75\%) & 3\\
        \hline
        Complexity & Model, measure and understand causal effects & 12 (75\%) & 3\\
        \hline
        Work structure and processes & Fixed structure of the organization & 11 (68,75\%) & 6\\
        \hline
        Resource conflicts & Costs and funding & 11 (68,75\%) & 6\\
        \hline
        Legal aspects & Data protection & 10 (62,5\%) & 8\\
        \hline
        People & Staff shortages & 10 (62,5\%) & 8\\
        \hline
        Systeme and Software & Interoperability & 9 (56,25\%) & 10\\
        \hline
        Data & Data sharing & 8 (50\%) & 11\\
        \hline
        People & Acquire competencies & 8 (50\%) & 11\\
        \hline
        SoS & Communication about the potential impacts of such systems & 8 (50\%) & 11\\
        \hline
    \end{tabular}
    \caption{Most cited challenges: challenges that were mentioned by half of the interviewees. The "mentions" column shows the number of interviewees that mentioned a given challenge and its corresponding percentage (in parenthesis). The "rank" column shows the overall ranking of each challenge (equal "mentions" values have the same rank).}
    \label{tab:challenges_most_cited}
\end{table}

\begin{table}[]
    \centering
    \begin{tabular}{|p{3cm}|p{6cm}|c|c|}
        \hline
        Category & Name & Mentions & Rank \\
        \hline
        Work structure and processes & Stakeholders with their own plans, goals and interests, who are not willing to cooperate with each other. & 12 (75\%) & 3\\
        \hline
        Work structure and processes & Fixed structure of the organization & 11 (68,75\%) & 6\\
        \hline
        Work structure and processes & Lengthy processes & 6 (37,5\%) & 17\\
        \hline
        Work structure and processes & High workload & 6 (37,5\%) & 17\\
        \hline
        Work structure and processes & Agreement and coordination between stakeholders takes too long & 5 (31,25\%) & 23\\
        \hline
        Work structure and processes & Transparency & 4 (25\%) & 31\\
        \hline
        Work structure and processes & Incompatible processes & 4 (25\%) & 31\\
        \hline
        Work structure and processes & Organizations are conservative & 2 (12,5\%) & 44\\
        \hline
    \end{tabular}
    \caption{Challenges from the category: work structure and processes. The "mentions" column shows the number of interviewees that mentioned a given challenge and its corresponding percentage (in parenthesis). The "rank" column shows the overall ranking of each challenge (equal "mentions" values have the same rank).}
    \label{tab:challenges_work_structure}
\end{table}

\begin{table}[]
    \centering
    \begin{tabular}{|p{3cm}|p{6cm}|c|c|}
        \hline
        Category & Name & Mentions & Rank \\
        \hline
        Data & Data silos & 14 (87,5\%) & 1\\
        \hline
        Data & Data capture & 12 (75\%) & 3\\
        \hline
        Data & Data sharing & 8 (50\%) & 11\\
        \hline
        Data & Data quality - accuracy & 7 (43,75\%) & 14\\
        \hline
        Data & Data quality - timeliness & 7 (43,75\%) & 14\\
        \hline
        Data & Data governance, data sovereignty & 6 (37,5\%) & 17\\
        \hline
        Data & Data quality - completeness & 5 (31,25\%) & 23\\
        \hline
        Data & Data quality - resolution & 5 (31,25\%) & 23\\
        \hline
        Data & Automation of data extraction, processing and use & 5 (31,25\%) & 23\\
        \hline
        Data & Data preparation & 4 (25\%) & 31\\
        \hline
        Data & High risk and uncertainty in the application of AI & 3 (18,75\%) & 39\\
        \hline
        Data & Data semantics & 2 (12,5\%) & 44\\
        \hline
    \end{tabular}
    \caption{Challenges from the category: data. The "mentions" column shows the number of interviewees that mentioned a given challenge and its corresponding percentage (in parenthesis). The "rank" column shows the overall ranking of each challenge (equal "mentions" values have the same rank).}
    \label{tab:challenges_data}
\end{table}

\begin{table}[]
    \centering
    \begin{tabular}{|p{3cm}|p{6cm}|c|c|}
        \hline
        Category & Name & Mentions & Rank \\
        \hline
        People & Communication between different actors or stakeholders. & 14 (87,5\%) & 1\\
        \hline
        People & Staff shortages & 10 (62,5\%) & 8\\
        \hline
        People & Acquire competencies & 8 (50\%) & 11\\
        \hline
        People & Change management & 6 (37,5\%) & 17\\
        \hline
        People & Technology affinity & 5 (31,25\%) & 23\\
        \hline
        People & Personal or political problems between different stakeholders & 5 (31,25\%) & 23\\
        \hline
        People & Network effect between stakeholders & 4 (25\%) & 31\\
        \hline
        People & Acceptance - open data, citizen data science & 4 (25\%) & 31\\
        \hline
        People & Acceptance - generic & 4 (25\%) & 31\\
        \hline
        People & Acceptance - fear of nature & 2 (12,5\%) & 44\\
        \hline
        People & Acceptance - fear of surveillance and loss of control & 2 (12,5\%) & 44\\
        \hline
        \end{tabular}
    \caption{Challenges from the category: people. The "mentions" column shows the number of interviewees that mentioned a given challenge and its corresponding percentage (in parenthesis). The "rank" column shows the overall ranking of each challenge (equal "mentions" values have the same rank).}
    \label{tab:challenges_people}
\end{table}

\begin{table}[]
    \centering
    \begin{tabular}{|p{3cm}|p{6cm}|c|c|}
        \hline
        Category & Name & Mentions & Rank \\
        \hline
        SoS & Communication about the potential impacts of such systems & 8 (50\%) & 11\\
        \hline
        SoS & Accountability, governance & 6 (37,5\%) & 17\\
        \hline
        SoS & Coupling of different models & 5 (31,25\%) & 23\\
        \hline
        SoS & Change in functionality & 4 (25\%) & 31\\
        \hline
        SoS & Security & 4 (25\%) & 31\\
        \hline
        SoS & Autonomous systems have to interact with living systems (high uncertainty) & 3 (18,75\%) & 39\\
        \hline
        SoS & Evaluation & 2 (12,5\%) & 44\\
        \hline
        SoS & Nature-based systems not considered first-citizen & 1 (6,25\%) & 52\\
        \hline
    \end{tabular}
    \caption{Challenges from the category: system-of-systems (SoS). The "mentions" column shows the number of interviewees that mentioned a given challenge and its corresponding percentage (in parenthesis). The "rank" column shows the overall ranking of each challenge (equal "mentions" values have the same rank).}
    \label{tab:challenges_sos}
\end{table}

\begin{table}[]
    \centering
    \begin{tabular}{|p{3cm}|p{6cm}|c|c|}
        \hline
        Category & Name & Mentions & Rank \\
        \hline
        Systeme and Software & Interoperability & 9 (56,25\%) & 10  \\
        \hline
        Systeme and Software & Robustness & 5 (31,25\%) & 23 \\
        \hline
        Systeme and Software & Missing software functionalities & 3 (18,75\%) & 39 \\
        \hline
        Systeme and Software & Usability & 3 (18,75\%) & 39 \\
        \hline
        Systeme and Software & Multifunctionality & 2 (12,5\%) & 44 \\
        \hline
        Systeme and Software & Speed & 2 (12,5\%) & 44 \\
        \hline
        Systeme and Software & Unused communication tools & 1 (6,25\%) & 52 \\
        \hline
        Systeme and Software & Computational effort for simulation & 1 (6,25\%) & 52 \\
        \hline
        Systeme and Software & Existing systems each have a single predefined and fixed function & 1 (6,25\%) & 52 \\
        \hline
        Systeme and Software & Existing hardware limitations & 1 (6,25\%) & 52 \\
        \hline
        Systeme and Software & Understanding & 1 (6,25\%) & 52 \\
        \hline
    \end{tabular}
    \caption{Challenges from the category: system and software. The "mentions" column shows the number of interviewees that mentioned a given challenge and its corresponding percentage (in parenthesis). The "rank" column shows the overall ranking of each challenge (equal "mentions" values have the same rank).}
    \label{tab:challenges_software}
\end{table}

\begin{table}[]
    \centering
    \begin{tabular}{|p{3cm}|p{6cm}|c|c|}
        \hline
        Category & Name & Mentions & Rank \\
        \hline
        Complexity & Model, measure and understand causal effects & 12 (75\%) & 3\\
        \hline
        Complexity & Unpredictability & 2 (12,5\%) & 44\\
        \hline
        Legal aspects & Data protection & 10 (62,5\%) & 8\\
        \hline
        Legal aspects & Different regulations need to be followed for SoS & 6 (37,5\%) & 17\\
        \hline
        Resource conflicts & Costs and funding & 11 (68,75\%) & 6\\
        \hline
        Resource conflicts & Conflict between different land uses & 7 (43,75\%) & 14\\
        \hline
        Resource conflicts & Nature-based systems have different needs over time & 3 (18,75\%) & 39\\
        \hline
    \end{tabular}
    \caption{Challenges from the categories: complexity, legal aspects, and resources conflics. The "mentions" column shows the number of interviewees that mentioned a given challenge and its corresponding percentage (in parenthesis). The "rank" column shows the overall ranking of each challenge (equal "mentions" values have the same rank).}
    \label{tab:challenges_other}
\end{table}


\section{Discussion}\label{section-discussion}
\subsection{Interpretation of the interview results}
First, the amount of challenges collected and the details of each category show the complexity of implementing an SoS in the smart city context.

The fact that challenges in categories "system and software" and "SoS" were ranked on average lower show us that challenges of technical nature, even if they do exist and need to be solved, are not the main pain points in implementing SoS in smart cities. What often came out of our discussions were that silos are a major challenge. These can be silos at the level of data: not knowing who owns what data and having difficulties to access it ("data silos", "data sharing" see table \ref{tab:challenges_data}); or silos at the level of processes and work structure: not being able to properly communicate of synchronize with other departments ("stakeholders with their own plans...", "fixed structure..." see table \ref{tab:challenges_work_structure}). These are challenges that are common to many AI or data-driven types of projects \cite{Reggio2020fail,Heidrich2022ailab} and Research and practical feedback in these types of projects could offer solutions (see for example the SE4AI domain \cite{Martinez2022survey}). Several interviewees mentioned that the application of agile methods could be a solution. Other challenges, that also happen in AI and data-driven projects, that were mentioned during the interviews are capturing data, pre-processing data, managing data quality and usage. The main cause of these challenges is the lack of skilled manpower, the costs of implementing such solutions, but also the very nature of data-driven projects: the potential of data might not be known in advance and different quality levels and resolution might be needed by different stakeholder.

\subsection{Comparison with challenges found in the related work}
We identified several papers in the related work that also described challenges related to SoS and smart cities. \cite{Axelsson2018} and \cite{Cavalcante2016,Cavalcante2017}, each identified eight challenges and opportunities related to mobility in Smart Cities. \cite{Mylonas2021} identified 11 challenges from the perspective of digital twins. 

The first aspect mentioned by these papers concerns the complexity related to the interoperability of different heterogeneous systems. \cite{Cavalcante2016,Cavalcante2017} identified  scale, that is, the number of devices being connected in cities, architecture, interoperability, and heterogeneity as main challenges and were also identified in our interviews. \cite{Mylonas2021} and \cite{Lenk2020} identified the need of standardization as one of the main issue when implementing a SoS. This was also a major concern from the interviewees.

All the authors recognized the huge potential of connecting the systems and make use of the data and information and identified challenges regarding the data \cite{Cavalcante2016,Cavalcante2017,Mylonas2021}. The ability to safely share information between systems is a challenge that is still not solved. In \cite{Cavalcante2016,Cavalcante2017}, the authors did not consider data quality issues raised by the interviewees and \cite{Mylonas2021}.
In addition, the amount of data available and the ability of systems and analytics to process it into useful and meaningful decisions is a concern that was cited by \cite{Cavalcante2016,Cavalcante2017} as well as our interviewees. The use of AI and its potential risks were only mentioned in our study.

\cite{Cavalcante2016,Cavalcante2017} and \cite{Westraadt2018} identified challenges related to the multitude of stakeholder, their disciplines and domains, for instance conflicts between stakeholders. This has also been confirmed with the conducted interviews. Work structures and processes have been identified as challenging. For a SoS project to work, processes need to be compatible, transparent and flexible which is often not possible due to the current stare structures of a city. Additionally, stakeholders need to be willing to adapt to new systems (socio-technical effects) \cite{Axelsson2018}.

The lack of communication and visualization tools can lead to misunderstandings, resulting in resistance to new technologies and new work processes, according to \cite{Westraadt2018}. As \cite{Lenk2020} mentioned, the accessibility to information systems and services for the various types of stakeholders need to be established to further increase the willingness. Furthermore trust of the users and new responsibilities were mentioned. In the end, communication not only between the systems but also between the different stakeholders are essential according to the interviewees and \cite{Mylonas2021}. These aspects were also identified in the interviews, but it was possible to address other issues related to social aspects, such as fears of staff shortages, lack of skills, lack of acceptance of open data and citizen data science, or the loss of control implied by the new technologies. The social aspects that need to be taken into account in cities are even more complex if we also consider the technical aspects. Another aspect indicated by \cite{Lenk2020} are the need for a strong political support and a cultural change, while \cite{Mylonas2021} focused on the development of skills to deal with the new technologies and processes.


\subsection{Sustainability}
Cities are still struggling to become more sustainable. While smart city strategies focus primarily on topics related to governance, infrastructure, and digital technologies, topics related to sustainability and especially the environment are rare. Even though sustainability is a key driver for the interviewees, they identified implementation as a major challenge due to the fixed structures of the city, their complexity, and the objectives of different stakeholders. The same observations were made by Joss et al. \cite{Joss}, also indicating that there are only a handful of smart cities that focus on the environment and sustainability, particularly green infrastructure.

According to our interviews, the lack of information regarding the structure and the land-use of a city to identify areas where greening would be possible hampers the development of a more sustainable city.

In addition, the complexity induced by the involvement of different stakeholders, their potentially conflicting interests and their current structures and work processes must be addressed to facilitate and find appropriate management strategies for sustainability. The complexity of the city must be understood and reduced in order to make decisions for greater sustainability. A structured holistic approach such as SoS engineering could be helpful.

One example is the project GreenTwins\footnote{\url{https://www.tallinn.ee/en/valisprojektid/greentwins-tallinn-helsinki-digital-green-model}} between the cities of Tallinn and Helsinki, which is being developed for a more democratic, resilient and green city. It takes into account several systems that influence green spaces in cities and the systems that are influenced by green spaces (urban climate, air pollution, human health, energy system, etc.) The green digital twin contributes to a better consideration of ecology in planning processes. The inclusion of a wide range of data allows for models and simulations so that the cities decision-making processes impact on sustainability can be better understood.

\section{Conclusions}\label{section-conclusions}
This paper presents and discusses the results of 23 interviews done between march and may 2022 within the context of the DYNASOS projects. We specifically focused on the use case of climate adaptation and greening in smart cities. We were able to collect 57 distinct challenges, to organize them, and to compare them with the one founds in the related work. Our findings suggest that while many technical issues remains (e.g., interoperability, data silos), the real blockers are non-technical (what \cite{Dogan2013,Ncube2018} referred as "human aspects" and \cite{Dridi2020} as "management and oversight challenges"): conflicts between stakeholders, inappropriate and inflexible work structures, etc. Challenges mentioned in the SoS literature such as the control of emergent phenomena, having to take into account multiple models are, for now, less of a concern for smart cities practitioners.
Finally, it is worth noting that many SoS development initiatives have been developed \cite{Lana2017} but it seems that very few of them were applied in the context of smart cities. Many of the people we spoke to commented that they don't know what the benefits of a SoS approach are, but the projects they are running and the problems they face are exactly those of system-of-systems. An evaluation of SoS development initiatives (such as the ones presented in \cite{Lana2017}) in the context of smart cities would greatly help practitioners

\bibliographystyle{alpha}
\bibliography{main}

\newcommand{\etalchar}[1]{$^{#1}$}
\begin{thebibliography}{MFBF{\etalchar{+}}22}

\bibitem[ABD15]{Albino2015}
Vito Albino, Umberto Berardi, and Rosa Dangelico.
\newblock Smart cities: Definitions, dimensions, performance, and initiatives.
\newblock {\em Journal of Urban Technology}, 22:2015, 02 2015.

\bibitem[AN18]{Axelsson2018}
J.~Axelsson and S.~Nylander.
\newblock An analysis of systems-of-systems opportunities and challenges
  related to mobility in smart cities.
\newblock pages 132--137, 2018.

\bibitem[BS06]{Boardman2006}
John Boardman and Brian Sauser.
\newblock System of systems-the meaning of of.
\newblock In {\em 2006 IEEE/SMC International Conference on System of Systems
  Engineering}, pages 6--pp. IEEE, 2006.

\bibitem[CCL{\etalchar{+}}16]{Cavalcante2016}
E.~Cavalcante, N.~Cacho, F.~Lopes, T.~Batista, and F.~Oquendo.
\newblock Thinking smart cities as systems-of-systems: A perspective study.
\newblock 2016.

\bibitem[CCLB17]{Cavalcante2017}
E.~Cavalcante, N.~Cacho, F.~Lopes, and T.~Batista.
\newblock Challenges to the development of smart city systems: A
  system-of-systems view.
\newblock pages 244--249, 2017.

\bibitem[DB08]{Dahmann2008}
Judith~S Dahmann and Kristen~J Baldwin.
\newblock Understanding the current state of us defense systems of systems and
  the implications for systems engineering.
\newblock In {\em 2008 2nd Annual IEEE Systems Conference}, pages 1--7. IEEE,
  2008.

\bibitem[DBB20]{Dridi2020}
C.~E. Dridi, Z.~Benzadri, and F.~Belala.
\newblock {System of Systems Modelling: Recent work Review and a Path Forward}.
\newblock {\em {ICAASE 2020 - Proceedings, 4th International Conference on
  Advanced Aspects of Software Engineering}}, 2020.

\bibitem[DeL05]{Delaurentis2005}
Daniel DeLaurentis.
\newblock Understanding transportation as a system-of-systems design problem.
\newblock In {\em 43rd AIAA aerospace sciences meeting and exhibit}, page 123,
  2005.

\bibitem[DL18]{Daneva2018}
M.~Daneva and B.~Lazarov.
\newblock Requirements for smart cities: Results from a systematic review of
  literature.
\newblock volume 2018-May, pages 1--6, 2018.

\bibitem[DNL{\etalchar{+}}13]{Dogan2013}
H.~Dogan, C.~Ncube, S.~L. Lim, M.~Henshaw, C.~Siemieniuch, M.~Sinclair,
  V.~Barot, S.~Henson, M.~Jamshidi, and D.~Delaurentis.
\newblock {Economic and societal significance of the systems of systems
  research agenda}.
\newblock {\em {Proceedings - 2013 IEEE International Conference on Systems,
  Man, and Cybernetics, SMC 2013}}, 2013.

\bibitem[EM21]{Elnashai2021}
A.~Elnashai and H.~Mahmoud.
\newblock A vision for smart and sustainable cities.
\newblock {\em IET Smart Cities}, 3(4):185--188, 2021.

\bibitem[GSB08]{Gorod2008}
Alex Gorod, Brian Sauser, and John Boardman.
\newblock System-of-systems engineering management: A review of modern history
  and a path forward.
\newblock {\em IEEE Systems Journal}, 2(4):484--499, 2008.

\bibitem[HHB{\etalchar{+}}13]{Henson2013}
Sharon~A Henson, MJD Henshaw, Vishal Barot, Carys~E Siemieniuch, Murray~A
  Sinclair, Mo~Jamshidi, Huseyin Dogan, SL~Lim, Cornelius Ncube, and Daniel
  DeLaurentis.
\newblock Towards a systems of systems engineering eu strategic research
  agenda.
\newblock In {\em 2013 8th International Conference on System of Systems
  Engineering}, pages 99--104. IEEE, 2013.

\bibitem[HJTV22]{Heidrich2022ailab}
Jens Heidrich, Andreas Jedlitschka, Adam Trendowicz, and Anna~Maria Vollmer.
\newblock Building ai innovation labs together with companies, 2022.

\bibitem[JSS{\etalchar{+}}19]{Joss}
Simon Joss, Frans Sengers, Daan Schraven, Federico Caprotti, and Youri Dayot.
\newblock The smart city as global discourse: Storylines and critical junctures
  across 27 cities.
\newblock {\em Journal of Urban Technology}, 26(1):3--34, 2019.

\bibitem[KKN{\etalchar{+}}22]{Koch2022}
Matthias Koch, Daniel Krohmer, Matthias Naab, Dominik Rost, and Marcus Trapp.
\newblock A matter of definition: Criteria for digital ecosystems.
\newblock {\em Digital Business}, 2(2):100027, 2022.

\bibitem[Len20]{Lenk2020}
U.~Lenk.
\newblock Smart cities and mbse: Comparison of concepts.
\newblock pages 169--174, 2020.

\bibitem[LSD{\etalchar{+}}17]{Lana2017}
C.~A. Lana, N.~M. Souza, M.~E. Delamaro, E.~Y. Nakagawa, F.~Oquendo, and J.~C.
  Maldonado.
\newblock {Systems-of-systems development: Initiatives, trends, and
  challenges}.
\newblock {\em {Proceedings of the 2016 42nd Latin American Computing
  Conference, CLEI 2016}}, 2017.

\bibitem[LTTG19]{Lee2019}
O.L. Lee, R.I. Tay, S.T. Too, and A.~Gorod.
\newblock A smart city transportation system of systems governance framework: A
  case study of singapore.
\newblock pages 37--42, 2019.

\bibitem[Mai98]{Maier1998}
Mark~W Maier.
\newblock Architecting principles for systems-of-systems.
\newblock {\em Systems Engineering: The Journal of the International Council on
  Systems Engineering}, 1(4):267--284, 1998.

\bibitem[MC19]{Monstadt2019}
J.~Monstadt and O.~Coutard.
\newblock Cities in an era of interfacing infrastructures: Politics and
  spatialities of the urban nexus.
\newblock {\em Urban Studies}, 56(11):2191--2206, 2019.

\bibitem[MFBF{\etalchar{+}}22]{Martinez2022survey}
Silverio Mart\'{\i}nez-Fern\'{a}ndez, Justus Bogner, Xavier Franch, Marc Oriol,
  Julien Siebert, Adam Trendowicz, Anna~Maria Vollmer, and Stefan Wagner.
\newblock Software engineering for ai-based systems: A survey.
\newblock {\em ACM Trans. Softw. Eng. Methodol.}, 31(2), apr 2022.

\bibitem[MKK{\etalchar{+}}21]{Mylonas2021}
G.~Mylonas, A.~Kalogeras, G.~Kalogeras, C.~Anagnostopoulos, C.~Alexakos, and
  L.~Munoz.
\newblock Digital twins from smart manufacturing to smart cities: A survey.
\newblock {\em IEEE Access}, 9:143222--143249, 2021.

\bibitem[NLA18]{Ncube2018}
C.~Ncube, S.~L. Lim, and {Amyot D., Maalej W., Ruhe G.}
\newblock {On systems of systems engineering: A requirements engineering
  perspective and research agenda}.
\newblock {\em {Proceedings - 2018 IEEE 26th International Requirements
  Engineering Conference, RE 2018}}, 2018.

\bibitem[PL20]{Prasetyo2020}
Y.A. Prasetyo and M.~Lubis.
\newblock Smart city architecture development methodology (scadm): A
  meta-analysis using soa-ea and sos approach.
\newblock {\em SAGE Open}, 10(2), 2020.

\bibitem[PLG20]{Payne2020}
B.~Payne, L.O. Ling, and A.~Gorod.
\newblock Towards a governance dashboard for smart cities initiatives: A system
  of systems approach.
\newblock pages 587--592, 2020.

\bibitem[RA20]{Reggio2020fail}
Gianna Reggio and Egidio Astesiano.
\newblock Big-data/analytics projects failure: A literature review.
\newblock In {\em 2020 46th Euromicro Conference on Software Engineering and
  Advanced Applications (SEAA)}, pages 246--255, 2020.

\bibitem[RAS22]{Adler2022}
Frank~Elberzhager Rasmus~Adler and Julien Siebert.
\newblock Towards a roadmap for trustworthy dynamic systems-of-systems.
\newblock paper accepted to the folowing conference: SERP'22 - The 20th Int'l
  Conf on Software Engineering Research and Practice (part of the 2022 World
  Congress in Computer Science, Computer Engineering, and Applied Computing
  (CSCE'22)), 2022.

\bibitem[WC18]{Westraadt2018}
L.~Westraadt and A.P. Calitz.
\newblock A gap analysis of new smart city solutions for integrated city
  planning and management.
\newblock pages 145--153, 2018.

\end{thebibliography}

\appendix
\section{Author contributions}
Both authors contributed equally to the manuscript, the analysis and the interview process.

\section{Funding}
This research was funded by the DynaSoS project (grant no. 01|S21104) of the German Federal Ministry of Education and Research (BMBF).

\section{Informed consent}
Informed consent was obtained from all subjects involved in the interviews.

\section{Conflicts of interest}
The authors declare no conflict of interest.
\end{document}